\documentclass[doublecol]{epl2} 
\usepackage{hyperref}

\title{Breakup dynamics of a neutron-halo projectile on heavy target at deep sub-barrier energies}
\shorttitle{Breakup dynamics of a neutron-halo projectile on heavy target at deep sub-barrier energies} 

\author{B. Mukeru\inst{1,2} \and T. Sithole \inst{1} \and 
Lauro Tomio\inst{3,4\footnote{Author to whom any correspondence should be addressed.}}}
\shortauthor{B. Mukeru \etal}

\institute{                    
\inst{1} Department of Physics, University of South Africa - Private Bag X6, Florida 1710, 
Johannesburg, South Africa\\
\inst{2} Universit\'e P\'edagogique Nationale, Av. de la Libération, P.O. Box 8815, 
Kinshasa, Democratic Republic of Congo\\
\inst{3} Centro Internacional de Física, Instituto de Física, Universidade de Brasília, 
70910-900, Brasília, DF, Brazil\\
\inst{4} Instituto de Física Teórica, Universidade Estadual Paulista, 
01140-070 São Paulo, SP, Brazil\\ \\
{\rm E-mail: mukerb1@unisa.ac.za,  sitholetapiwa81@gmail.com, and lauro.tomio@unesp.br}
}
\abstract{
    By studying the total fusion and breakup cross-sections in the interaction of
    the neutron-halo $^{11}{\rm Be}$ projectile on the lead target $^{208}$Pb,
  it is shown that, even for the neutron-halo projectile, the breakup channel
  remains the most dominant reaction channel at sub-barrier energies, 
 following a characteristic behavior that was also previously verified
  for the case of the proton-halo projectile $^8{\rm B}$. 
  This feature is found to emanate from the enhancement of the 
  breakup cross-section, due to the continuum-continuum 
  couplings coming exclusively from its Coulomb component. 
  We further speculate that the 
  enhancement of the Coulomb breakup cross-section at sub-barrier incident energies 
  by the continuum-continuum couplings
  could be associated with the projectile breaking up on the outgoing trajectory, provided 
  these couplings can be proven to delay the breakup process.}

\begin{document}

\maketitle

\section{Introduction}
A recent experimental measurement of the breakup of $^8{\rm B}$ proton-halo nucleus on 
a lead target at deep sub-barrier energies by Pakou \textit{et al}.\cite{Pakou2020},
yielded a quite interesting result: 
the breakup channel is reported to be the main reaction channel at these energies. 
Intuitively, one assumes that at deep sub-barrier energies, the reaction should be 
dominated by reaction channels other than the breakup channel. In a subsequent 
study~\cite{MukeruPR2021}, it was evidenced that the predominance of the breakup channel
over the fusion channel in this incident energy region could be attributed to couplings 
among the continuum states of the projectile. Using the continuum discretized coupled 
channels (CDCC) calculations, further analysis led to the conclusion that such 
continuum-continuum couplings (CCC) indicate the breakup on the outgoing trajectory. 
While these couplings are known to suppress strongly the breakup cross-sections at
incident energies above the Coulomb barrier
\cite{Nunes1999,Summers2004,Canto2009,Diaz2002,Mukeru2015,MukeruJPG2018,
Mukeru2018,MukeruCPC2022}, 
it was verified in Ref.\cite{MukeruPR2021} that they enhance the breakup cross-section
at sub-barrier energies, being further argued that this enhancement could be attributed 
to the breakup occurring on the outgoing trajectory. However, the question remains 
open whether the projectile breaks up on its incoming trajectory toward the target or 
on its outgoing trajectory as it leaves the target and how that affects the breakup dynamics.

A subsequent analysis in Ref.~\cite{Yang2022}, of the same reaction with the $^8{\rm B}$
proton-halo nucleus, within the same incident energy range, further confirmed the
findings of Ref.~\cite{Pakou2020}, by indicating the effect of Coulomb polarization on
the proton halo state, with the correlation information revealing that the prompt 
breakup mechanism dominates, occurring predominantly on the outgoing trajectory. This
assertion corroborates the conclusion anticipated in Ref.~\cite{MukeruPR2021} that the
breakup of the projectile occurs in the outgoing trajectory.

Particularly, in Ref.\cite{Yang2022}, it was emphasized the relevance in elucidating 
the long-standing question about the breakup dynamics of a proton halo nucleus. 
Relying on the results obtained near Coulomb barrier energies, their analyses have 
signaled distinctive dynamics of a proton-halo nucleus as compared with a neutron-halo
nucleus, which has been assigned to the Coulomb effect of the halo proton, 
indicating little influence of the continuum on elastic scattering and complete fusion. 
Nevertheless, as commented in Ref.~\cite{Yang2022}, further investigations are 
still desired to elucidate the breakup behavior of a proton halo nuclear system, 
which can be quite relevant in the light of potential astrophysical implications. 
In this regard, one could check whether the conclusions reported in 
Refs.~\cite{Pakou2020,MukeruPR2021,Yang2022} can be assumed as a universal
signature of the breakup of weakly-bound systems at deep sub-barrier energies, 
by first extending the same investigation to neutron-halo projectiles. 
The fundamental difference between proton-halo and neutron-halo is the presence 
of the Coulomb barrier in the core-proton system, which is absent in the 
core-neutron system since the neutron is not charged. Therefore, considering 
a neutron-halo projectile for the same study would provide an opportunity to 
test, whether the importance of the breakup channel over other reaction channels 
at deep sub-barrier incident energies emanate from dynamical effects 
(associated with the projectile-target interaction), or from static effects 
(associated with the projectile ground-state wave function).
On the competition between the breakup and fusion channels below the Coulomb barrier, 
it was also shown previously that the breakup cross-section becomes dominant over 
the fusion cross-section, in Refs.~\cite{MukeruGR2015,Otamar2013}. By assuming the 
same heavy target $^{208}$Pb, this was shown in Ref.~\cite{MukeruGR2015}, when 
studying the Coulomb barrier penetrability using the $^{11}$Be projectile; and, 
in Ref.~\cite{Otamar2013}, by considering the $^6$Li as the projectile, treated 
as a weakly-bound cluster of an alpha particle with a deuteron. In short, despite 
the registered successes over the past decades in probing nuclear reactions by 
weakly-bound exotic projectiles in heavy targets, which can be traced through 
recent studies~\cite{2020Lha,Rangel2020,Canto2020,MukeruPRC2020,Wang2021,
Pietro2023,Ferreira2023,MukeruEPL2023,MukeruPRC2023}, with their cited references,
these investigations  at deep sub-barrier energies are still limited, as most 
available results in this topic are based on incident energies above and around 
the Coulomb barrier.

Our aim in this paper is to report an extension of the mentioned studies on 
proton-halo reactions~\cite{Pakou2020,MukeruPR2021,Yang2022}, by considering a 
neutron-halo projectile, to probe the possible universality of previous conclusions. 
To this end, we study the breakup dynamics of the $s-$wave neutron-halo nucleus 
$^{11}{\rm Be}$ on a lead target at Coulomb sub-barrier and around-barrier incident 
energies. The main goal is to verify whether, for deep sub-barrier incident energies, 
the breakup cross-section remains dominant over the total fusion cross-section as 
in the case of proton-halo projectiles, such that some universal characteristics 
can be extracted.

\section{Brief theoretical approach} 
The fundamental mathematical formulation of the CDCC (continuum discretized coupled channels) 
formalism, which is the theoretical approach used in this work, can be found in 
Ref.~\cite{Austern1987}, such that we avoid giving more details here. For a more recent review 
of CDCC with its theoretical foundation, we have also the Ref.~\cite{2012Yahiro}.
Within the CDCC formalism, once the total wave function is expanded on the projectile bound and 
bin states (whose wave functions are square-integrable), a truncated set of coupled differential 
equations of the radial part $\chi_\beta(R)$ of the wave function is obtained, which contains the 
coupling matrix elements 
{\small
\begin{eqnarray}\label{CME}
U_{\beta\beta^\prime}^{LL^\prime J}(R) = \langle\mathcal{Y}_\beta^{LJ}(\bm{r,\Omega_R})
|U_{pt}(\bm{r,R})|\mathcal{Y}_{\beta^\prime}^{L^\prime J}(\bm{r,\Omega_R})\rangle,
\end{eqnarray}
}where $\mathcal{Y}_\beta^{LJ}(\bm{r,\Omega_R})$, is the channel wave function that 
contains bound and bin wave functions, $\Omega_{\bm R}$ is the solid angle in the direction of 
the projectile-target center-of-mass $\bm R$, expressed in spherical coordinates, with $L$ and 
$J$ being the orbital and total angular momentum quantum numbers. In Eq.~(\ref{CME}),
$U_{pt}(\bm{r,R}) = U_{ct}(\bm R_{ct}) +U_{vt}(\bm R_{vt})$, with $U_{ct}$, and $U_{vt}$, 
are core-target and valence-nucleon-target optical potentials, having the corresponding coordinates 
\begin{eqnarray}\label{coord}
{\bm R_{ct}} \equiv {\bm R} + \frac{1}{A_p}{\bm r},\quad
{\bm R_{vt}} \equiv {\bm R} - \frac{A_c}{A_p}{\bm r},
\end{eqnarray}
where $A_c$ and $A_p = A_c+1$, are the core and projectile atomic mass numbers, respectively. In 
Eq.~(\ref{CME}), $\beta\equiv (\alpha_b,\alpha_i)$, where $\alpha_b$ represents the quantum 
numbers that describe the projectile bound state, with $\alpha_i$ standing for the quantum 
numbers that describe the bin states, with $i=1,2,\ldots, N_b$, where $N_b$ is the number of bins. 

The coupling matrix elements (\ref{CME}), can be split into couplings to and from the 
bound-state $U_{\alpha_b\alpha}^{L J}(R)$, and couplings among the continuum states 
$U_{\alpha_i\alpha_i^\prime}^{LL^\prime J}(R)$, given by
{\small\begin{eqnarray}\label{DC}
U_{\alpha_b\alpha_i}^{L J}(R) & = & 
\langle\mathcal{Y}_{\alpha_b}^{LJ}(\bm{r,\Omega_R})
|U_{pt}(\bm{r,R})|\mathcal{Y}_{\alpha_i}^{L^\prime J}(\bm{r,\Omega_R})\rangle,\nonumber\\
U_{\alpha_i\alpha_i^\prime}^{LL^\prime J}(R) & = &
\langle\mathcal{Y}_{\alpha_i}^{LJ}(\bm{r,\Omega_R})
|U_{pt}(\bm{r,R})|\mathcal{Y}_{\alpha_i^\prime}^{L^\prime J}(\bm{r,\Omega_R}).
\end{eqnarray}
}The coupling matrix elements are evaluated subject to the boundary conditions 
in the asymptotic region ($R\to\infty$),
{\small
\begin{eqnarray}\label{BC}
\chi_{\beta}(R)\stackrel{R\to\infty}\to\frac{\rm i}{2}\left[H_{\alpha_i}^-(K_{\alpha_i}R)
	\delta_{\alpha_b\alpha_i}-H_{\alpha_i}^+(K_{\alpha_i}R)
	S_{\beta\beta^{\prime}}^J(K_{{\beta}^{\prime}})\right],\!\!\!
\end{eqnarray}
}where $H_{\beta}^{\pm}(K_{\beta}R)$ are Coulomb-Hankel functions \cite{Thompson1988}, and
$S_{\alpha_i\alpha_i^{\prime}}^J(K_{\alpha_i^\prime})$ are the scattering S-matrix elements, 
with $K_{\alpha_i} = \sqrt{\frac{2\mu_{pt}(E-\varepsilon_{\alpha_i})}{\hbar^2}}$, where 
$\mu_{pt} = m_0A_pA_t/(A_p+A_t)$
($m_0$ is the nucleon's mass and $A_t$ the projectile atomic mass number) 
is the projectile-target reduced mass,  $E$ is the incident energy, with 
$\varepsilon_{\alpha_i}$ being the bin energies.

The breakup cross-section can be directly obtained from the scattering matrix as follows 
\cite{Pierre2017}
\begin{eqnarray}\label{BCS}
\sigma_{\rm BU} = \frac{\pi}{K_{\alpha_b}^2}\sum_{J\alpha_iL\alpha_i^\prime L^\prime}
\frac{2J+1}{2j+1}|S_{\alpha_i\alpha_i^{\prime}}^J(K_{\alpha_i^\prime})|^2,
\end{eqnarray}
where $j$ is the total angular momentum associated with the core-nucleon relative motion,
and $K_{\alpha_b}$, is the initial relative momentum, which is related to the final 
relative momentum $K_{\alpha_i}$ through the following energy conservation equation
$\frac{\hbar^2 K^2_{\alpha_i}}{2\mu_{pt}} -\varepsilon_{\alpha_i} = \frac{\hbar^2 
K^2_{\alpha_b}}{2\mu_{pt}} +\varepsilon_b$ (where $\varepsilon_b<0$ is the 
ground-state binding energy). 
The total fusion cross-section can be obtained as follows:
{\small
\begin{eqnarray}\label{TF}
  \sigma_{\rm TF} & = & \sum_{J = 0}^{J_{max}}\sigma_{\rm TF}^{(J)} \\
        \sigma_{\rm TF}^{(J)} & = &  \frac{2\mu_{pt}}{\hbar^2K_{\alpha_b}}(2J+1)
        \sum_{\beta\beta_i}\chi^{LJ}_{\beta}(R)
        |W^{LL^\prime J}_{\beta\beta^\prime}(R)|\chi^{L^\prime J}_{\beta}(R),\nonumber
\end{eqnarray}
}where $W^{LL^\prime J}_{\beta\beta^\prime}(R)$ are the imaginary parts of the coupling 
matrix elements $V^{LL^\prime J}_{\beta\beta^\prime}(R)$
that contain the imaginary parts of the potential $U_{pt}(\bm{r,R})$. Therefore, they are 
responsible for nuclear absorption.

\subsection{Projectile-target potentials}

A selection of the projectile-target potentials necessary to calculate both fusion and 
breakup cross-sections on the same
footing can prove to be a challenging task. The main reason is the fact that both 
cross-sections emanate from different dynamics. Quite often in the literature, the 
Woods-Saxon form factor is used to model the real and imaginary parts of the
potentials $U_{pt}(\bm{r,R})$. With the coordinate definitions (\ref{coord}) and
$n\equiv ct,vt$, 
\begin{eqnarray}\label{pot}
  U_n(\bm R_n) & = & V_n(\bm R_n)+{\rm i}W_n(\bm R_n)\nonumber\\
  & = & \frac{V_{0}^{(n)}}{1+\exp[({\bm R}_n-{\bm R}_{0}^{(n)})/a_{0}^{(n)}]} \\
  & + & \frac{ {\rm i}\; W_{0}^{(n)}}{1+\exp[(\bm R_n-R_{w}^{(n)})/a_{w}^{(n)}]},\quad n\equiv ct,vt \nonumber
\end{eqnarray}
where $V_{0}^{(n)}$ and $W_{0}^{(n)}$ are the depths of the real and imaginary parts, respectively, 
$R_{0}^{(n)} = r_{0}^{(n)}(A_n^{1/3} + A_t^{1/3})$ and $R_{w}^{(n)} = r_{w}^{(n)}(A_n^{1/3} + A_t^{1/3})$ are 
the corresponding nuclear radii, with $a_{0n}$ and $a_{wn}$ the respective diffuseness.
The potentials (\ref{pot}) are used in the off-diagonal channels to couple bound to 
continuum and continuum to continuum channels. In the elastic scattering channel, 
the real and imaginary parts of the optical potential represent the expected value 
of $U_{pt}(\bm{r,R})$, concerning the ground state of the projectile nucleus:
\begin{eqnarray}\label{Dpot}
  V_{\alpha_b\alpha_b}(\bm{R}) & = & \int d^3\bm r|\phi_{\alpha_b}(\bm r)|^2V_{pt}(\bm{r,R}),
  \nonumber\\
  W_{\alpha_b\alpha_b}(\bm{R}) & = & \int d^3\bm r|\phi_{\alpha_b}(\bm r)|^2W_{pt}(\bm{r,R}),
\end{eqnarray}
where $\phi_{\alpha_b}(\bm r)$ is the ground state wave function, and $V_{pt}(\bm{r,R}) = 
V_{ct}(\bm R_{ct}) + V_{vt}(\bm R_{vt})$,
$W_{pt}(\bm{r,R}) = W_{ct}(\bm R_{ct}) + W_{vt}(\bm R_{vt})$, are the real and imaginary parts 
of $U_{pt}(\bm{r,R})$. 
Given the longer tail of the projectile nucleus, due to its low breakup threshold, the nuclear forces 
are extended well beyond the barrier radius through the tails of $V_{pt}(\bm{r,R})$ and $W_{pt}(\bm{r,R})$.
As such, $V_{\alpha_b\alpha_b}(\bm{R})$ will result in lowering the Coulomb barrier, whereas 
$W_{\alpha_b\alpha_b}(\bm{R})$ will exhibit a long-range absorption behavior. Consequently, 
the total fusion obtained with these potentials is expected to be much larger. 
 Realistic fusion cross-sections (i.e., comparable with the experimental data) are generally
  obtained by considering short-range imaginary potentials. For example, in
Refs.\cite{Hagino2000,Diaz2002,Ferreira2023,Lubian2022,Camacho2018},
 strong short-range $W_{ct}$, $W_{vt}$, and $W_{pt}$, with parameters 
$W_0 = 50$\,MeV, $r_w = 1.0$ and $a_w = 0.1$\,fm, are adopted.
However, such a choice of imaginary potentials may not be suitable in
the calculations of breakup cross-sections.

\section{Details of the numerical calculations}

Here we provide the necessary information on the breakup dynamics, by describing the internal structure 
of the neutron-halo projectile nucleus $^{11}$Be. 
It is modeled as a $^{10}$Be core nucleus to which a valence neutron is weakly bound in the following 
$s-$wave configuration $^{10}{\rm Be}\,\otimes\,n(2s_{\frac{1}{2}^+})$, with $\ell_0=0$,
where $\ell_0$ is the ground-state orbital angular momentum associated with the core-neutron relative motion.
The binding energy of this ground state is $\varepsilon_0=-0.504$\,MeV \cite{Wang2017}. 
Also, this nucleus exhibits a first excited bound-state with energy $\varepsilon_1=-0.183$\,MeV 
in the $p_{\frac{1}{2}^-}$ state ($\ell_0=1$), and a narrow resonance with energy 
$\varepsilon_{res}=1.274$\,MeV, in the $d_{\frac{5}{2}^+}$ continuum state.

To obtain the internal states of the $^{11}$Be nucleus (i.,e., bound and scattering states), the two-body 
Schr\"odinger equation is numerically solved, using the Woods-Saxon potential with both central
and spin-orbit coupling components. For the numerical values of the different parameters, we assume the same
ones considered in Ref.\cite{MukeruCPC2022}, which were taken from Ref.~\cite{Capel2003}.
  As already indicated, it is not straightforward the procedure in determining which common
   imaginary potentials to use in simultaneous calculations of both cross-sections. 
 In this study, the choice of the long-range imaginary potentials was motivated by the
  understanding that these potentials are expected to provide realistic calculations
  for the breakup cross-section, but end up overestimating the total fusion cross-section.
  Therefore, once the breakup cross-section is found to be larger than the total fusion
  counterpart, this would be even more so when short-range imaginary potentials are
  used. This happens because short-range imaginary potentials are expected to
 enhance the breakup cross-section. 
 By taken from Ref.~\cite{Capel2003},
the real and 
imaginary parts of the core-target optical potential parameters used in the construction
of the projectile-target coupling matrix elements, taken from Ref.~\cite{Capel2003}, are 
$V_0=70$\,MeV, $R_0=7.43$\,MeV, $a_0=1.04$\,MeV, $W_0 = 58.9$\,MeV, $R_w = 7.19$\,MeV, and $a_w = 1.0$\,MeV.
For the  neutron-target optical potential, the  global parametrization of 
Ref.~\cite{Becchetti1969}, was adopted. 
These potentials, together with
  the folding potential (\ref{Dpot}), in the elastic scattering channel, 
  extend the absorption to outside the usual region, increasing the fusion
  cross-section.
  So, we need to be mindful of the choice of imaginary potentials in
  the analysis of the results.
 To obtain
fusion cross-sections that are comparable with the available experimental data
(as it happens for $^{11}{\rm B}+{}^{209}{\rm Bi}$ \cite{Hinde2010}),
and test how the total fusion-cross section is overestimated of the
  long-range imaginary potentials, we will perform another set of calculations where
we replace the long-range
  $W_{ct}$, $W_{nt}$ and $W_{pt}$ by the short-range ones, i.e.,
  $W_0 = 50$\,Mev, $r_w = 1.0$\,MeV and $a_W = 0.1$\,MeV, as in Refs.
\cite{Hagino2000,Diaz2002,Ferreira2023,Lubian2022,Camacho2018,Signorini2004}.
  However, by calculating with long- or short-range imaginary potential, our interest is to verify whether the breakup 
  cross-section remains larger than the fusion cross-section
at incident energies below the Coulomb barrier.
And, if that is the case, why this happens.

For solving the coupled differential equations emanating from the projectile-target three-body 
Schr\"odinger equation, various numerical parameters were optimized to satisfy the convergence 
requirements. In this regard, the following maximum limiting values were applied:
For the core-neutron, the orbital angular momentum $\ell$ was truncated at $\ell_{max}=6\hbar$,
with $r_{max}=100$\,fm being the maximum radial coordinate $r$, and 
$\varepsilon_{max}=8$\,MeV the maximum excitation energies $\varepsilon$.
For the projectile target, we have $L_{max}=1000\hbar$ and $R_{max}=500$\,fm, respectively,
for the maximum orbital angular momentum $L$ and for the radial coordinate $R$. 
Also, the radial coordinates $r_{max}$ and $R_{max}$ were sliced into radial mesh points equally 
spaced by $\Delta r = 0.1$\,fm and $\Delta R = 0.05$\,fm, respectively.
The projectile-target potentials were expanded into potential multipoles up to $\lambda_{max}=4$.  
The energy interval
$[0,\varepsilon_{max}]$ was discretized into energy bins of widths,  
$\Delta\varepsilon=0.5\,{\rm MeV}$, for the
$s$- and $p$-states; $\Delta\varepsilon=1.0\,{\rm MeV}$, for the $f$- and 
$d$-states; $\Delta\varepsilon=1.5\,{\rm MeV}$,
for ${\rm g}$-states; and $\Delta\varepsilon=2.0\,{\rm MeV}$, for higher partial waves.
Finer bins were considered for the resonant state. The 
numerical calculations were performed with FRESCO computer codes \cite{Thompson1988}.
\begin{figure}[!!t]
	\begin{center}
		\resizebox{85mm}{!}{\includegraphics{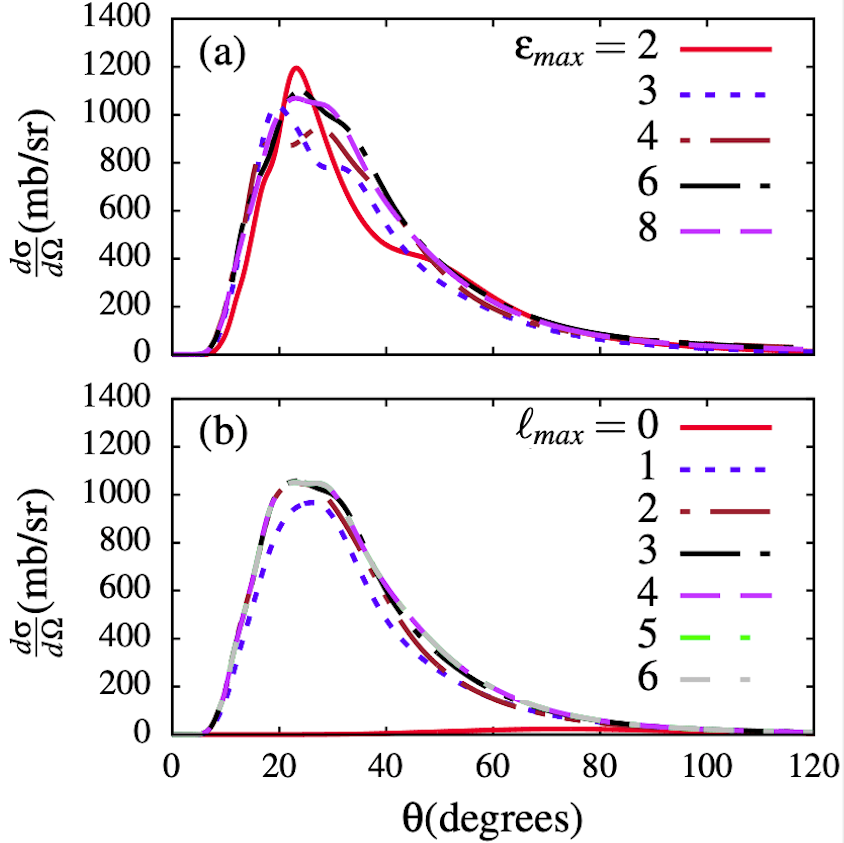}}
	\end{center}
	\caption{\label{f01} Convergence results for the differential breakup 
 cross-sections in terms of the maximum core-neutron excitation energies $\varepsilon_{max}$ 
 (MeV units) [panel (a)] and maximum core-neutron orbital angular momentum $\ell_{max}$ 
 ($\hbar$ units) [panel (b)].}
\end{figure}
In Fig.~\ref{f01}, we show samples of convergence tests in 
    terms of $\ell_{max}$ and $\varepsilon_{max}$ for $E_{cm}/V_{\rm B}=0.8$,
      where $V_{\rm B}=37.90$\,MeV is obtained from the S\~ao Paulo potential 
      (SPP) \cite{Chamon2002}. As one can verify from the given results in this figure, 
      the convergence is well reached
      for $\varepsilon_{max} = 8$\,MeV, and $\ell_{max} = 6\hbar$.

\section{Results and Discussion}
\begin{figure}[!!t]
	\begin{center}
		\resizebox{85mm}{!}{\includegraphics{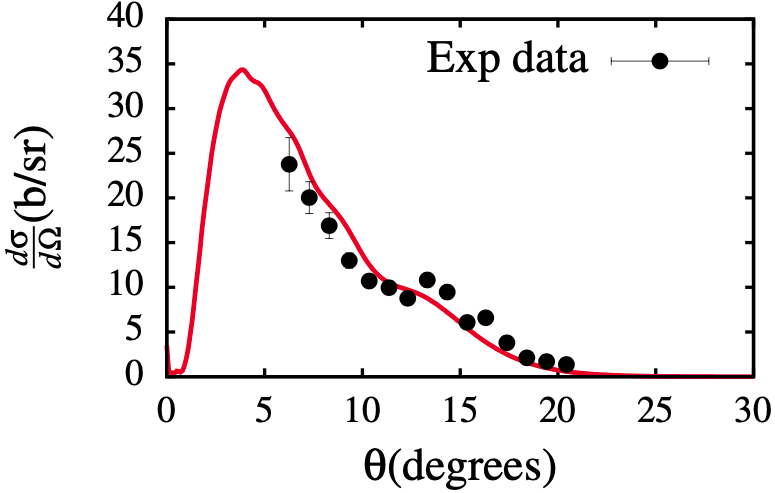}}
	\end{center}
	\caption{\label{f02} Comparison of the theoretical computed breakup cross-section
 (solid line) with the corresponding experimental data, measured at $E_{lab}=140$\,MeV, 
 as given in Ref.~\cite{Duan2020}.}
        \end{figure}

To further assess the stability of the numerical results and how well the experimental
data can be described, we first compare in Fig.\ref{f02},
the computed results for the differential breakup cross-section with the available experimental data,
measured at $E_{lab}=140$\,MeV, as given in Ref.~\cite{Duan2020}. 
As shown, the experimental data are quite well reproduced.
We also found that (results not shown) the breakup cross-section with the short-range 
imaginary potential becomes highly oscillatory, particularly at small angles, 
and does not provide a good fit for the experimental data as the ones obtained
with the long-range imaginary potential. This is why, as already pointed out, 
we chose to use the long-range imaginary potential to calculate the breakup 
cross-sections.

\begin{figure}[!!t]
	\begin{center}
		\resizebox{85mm}{!}{\includegraphics{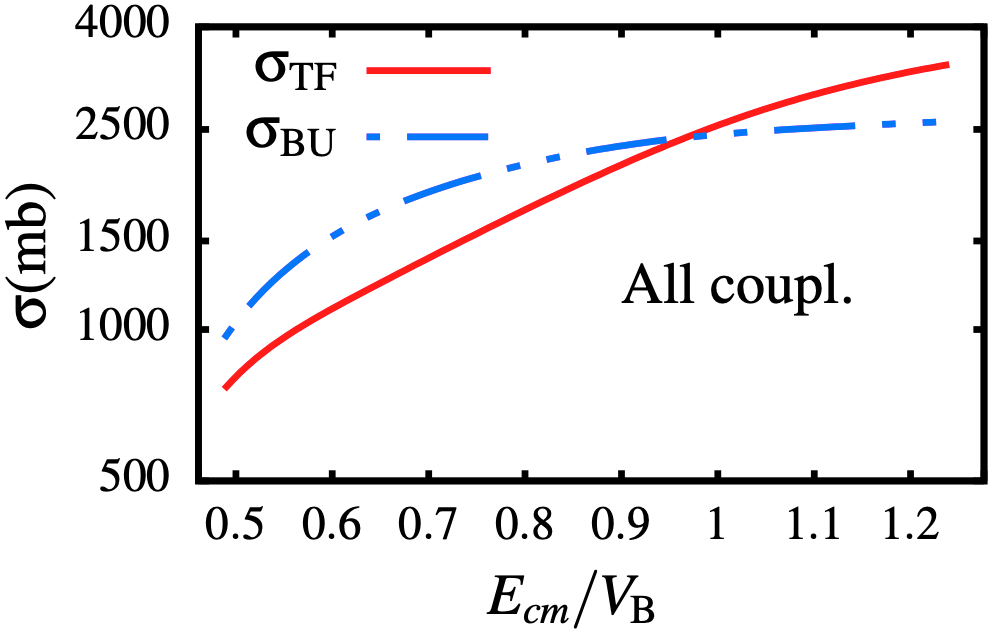}}
	\end{center}
	\caption{\label{f03} Breakup (dash-dotted line) and total fusion (solid line) cross-sections,
 plotted as functions of the incident
          energy scaled by the Coulomb barrier height $V_{\rm B}$, and obtained when all the different 
          coupled channels are included in the coupling matrix elements.}
\end{figure}

Figure \ref{f03} shows the breakup and total fusion cross-sections, respectively $\sigma_{\rm BU}$ 
and $\sigma_{\rm TF}$, as functions of the ratio between total energy $E_{cm}$ and potential 
barrier height $V_B$,  with $E_{cm}/V_{\rm B}$ in the interval $[0.5,1.3]$, 
The same potential parameters were applied in both calculations to obtain the 
breakup cross-section in Fig.\ref{f02}. Although these parameters increase the
total fusion cross-section, both fusion and breakup cross-sections are treated on the same 
footing with this choice, with the outcome not affected by different calculations. 
The results in Fig.\ref{f03} were obtained with all the different 
couplings being included in the coupling matrix elements, identified as ``All coupl.", i.e., 
with couplings to and from the projectile bound-state and continuum-continuum couplings.
One observes that at sub-barrier energies ($E_{cm}/V_{\rm B}\le 1$), the breakup cross-section is
dominant over the total fusion cross-section. The transition occurs around the Coulomb barrier 
where the total fusion cross-section prevails. Therefore, one can infer that, even in the case of 
a neutron-halo weakly-bound projectile, the breakup channel remains the dominant reaction channel 
at sub-barrier incident energies.  It is interesting to see that even when long-range 
imaginary potentials are considered which are known to increase the fusion cross-section, 
the breakup cross-section remains more important than its fusion counterpart.
In this regard, it follows that the conclusions of Refs.~\cite{Pakou2020,MukeruPR2021,Yang2022} 
on proton-halo projectile can also be extended to a neutron-halo projectile.
We do not expect the use of short-range imaginary potentials
  to reverse this trend at sub-barrier energies, but such potentials can be expected to push
  the transition point where the fusion cross-section becomes more important for larger
incident energies.
The results in Fig.\ref{f03}, further suggests that the Coulomb barrier in
the core-proton system is not
responsible for the importance of the breakup cross-section over the total fusion 
cross-section at sub-barrier incident energies, which implies that static effects
related to the 
projectile ground-state wave function are not the main factor contributing
to this phenomenon.  
As argued in Ref.~\cite{MukeruPR2021}, this leaves dynamical effects (due to the projectile-target 
interaction) as one of the main factors responsible for the importance of the breakup
cross-section over
the total fusion cross-sections at incident sub-barrier energies.
A similar trend regarding the importance of the breakup cross-section over the total fusion cross-section
was also reported in the breakup of $^6$Li nucleus (treated as a weakly-bound cluster of an alpha particle 
and the deuteron) on the same target nucleus~\cite{Otamar2013}. 

\begin{figure}[!!t]
	\begin{center}
  \resizebox{85mm}{!}{\includegraphics{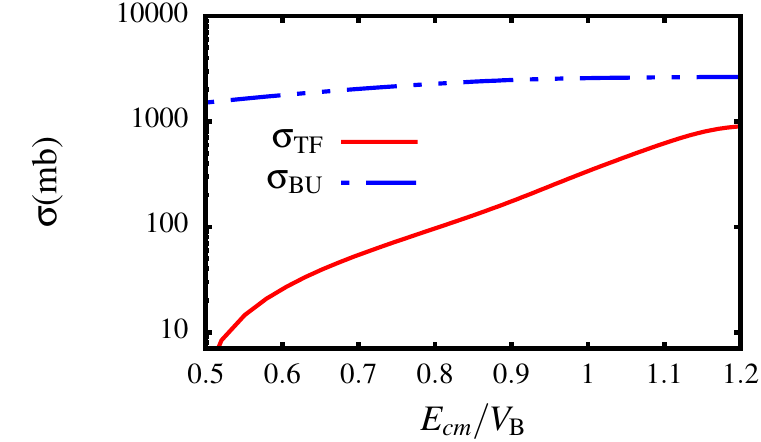}}
	\end{center}
	\caption{\label{f04}
 Fusion and breakup cross-section as a 
 function of $E_{cm}/V_{\rm B}$, obtained when the long-range imaginary 
 part of the nuclear potential is replaced by the short-range one, i.e., with
 $W_0 = 50$\,Mev, $r_w = 1.0$\,MeV and $a_w = 0.1$\,MeV for 
 core-target, neutron-target and projectile-target imaginary potentials.}
        \end{figure}

 To verify how the long-range imaginary potentials overestimate
  the total fusion cross-section, we repeated the same calculations, but with the
  short-range imaginary potentials, as described in the previous section.
  The calculated fusion and breakup cross-sections are shown in Fig.\ref{f04}.
  Compared to Fig.\ref{f03}, it is noticed that the total fusion 
  cross-sections are largely suppressed, and the breakup cross-section becomes 
  substantially larger than the fusion cross-section. Although the fusion
  cross-sections, as shown in Fig.\ref{f04}, are not considered in the following discussion,
  we are aware of this suppression. 
Within a careful comparison, the Figs.\ref{f03} and \ref{f04} suggest that 
the difference between both breakup cross-sections is not very pronounced. However, one can verify that 
the results obtained with short-range imaginary potentials 
are significantly larger than those obtained with long-range imaginary potentials.
  
\begin{figure}[!!h]
	\begin{center}
			\resizebox{85mm}{!}{\includegraphics{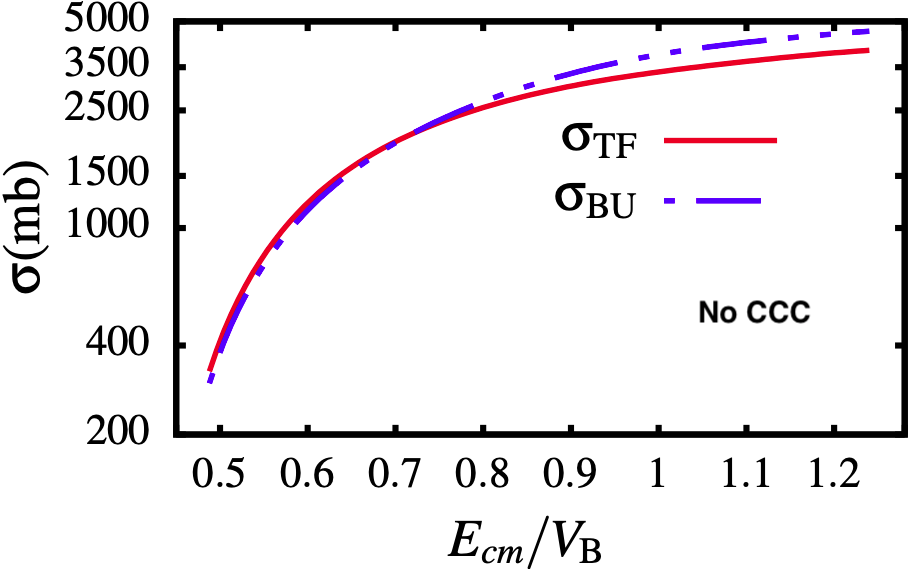}}
	\end{center}
	\caption{\label{f05}Breakup and total fusion cross-sections as functions of the incident energy
          scaled by the Coulomb barrier height $V_{\rm B}$, and obtained when the continuum-continuum couplings
          are excluded (``No CCC'') from the coupling matrix elements.}
\end{figure}

To gain more insight into the importance of the breakup cross-section over the total fusion cross-section
at sub-barrier energies, it is essential to investigate the effect of the continuum-continuum couplings.
In Fig.\ref{f05}, we compare the total fusion and breakup cross-sections obtained when the 
continuum-continuum couplings are removed from the coupling matrix elements (without CCC), i.e., 
leaving a single transition to and from the projectile bound state.
By inspecting the results shown in that figure  a stark difference from Fig. \ref{f03} is observed  (with all couplings),
as one notes that, at sub-barrier energies
($E_{cm}/V_{\rm B}\le 0.8$), both breakup and total fusion cross-sections are almost similar, 
with the two curves becoming hardly distinguishable. 
 Above the Coulomb barrier, the breakup cross-section starts becoming dominant. So, by comparing Figs.\ref{f03} and \ref{f05}, the continuum-continuum couplings appear to be responsible
for the quantitative importance of the breakup cross-section over the fusion cross-section at sub-barrier incident energies,
as also previously reported in Ref.\cite{MukeruPR2021}.
 However, considering the fusion calculations as given 
in Fig.\ref{f04}, with short-range imaginary part in the nuclear potential, 
the fusion cross-section in the absence of the continuum-continuum
couplings would be lower than the breakup cross-section even at sub-barrier energies.
Therefore, although Fig.~\ref{f05} does not provide a realistic picture, 
it points out the fact that when the continuum-continuum couplings are removed 
from the matrix elements, the gap between the two curves will be 
significantly narrowed.

\begin{figure}[!!t]
	\begin{center}
			\resizebox{85mm}{!}{\includegraphics{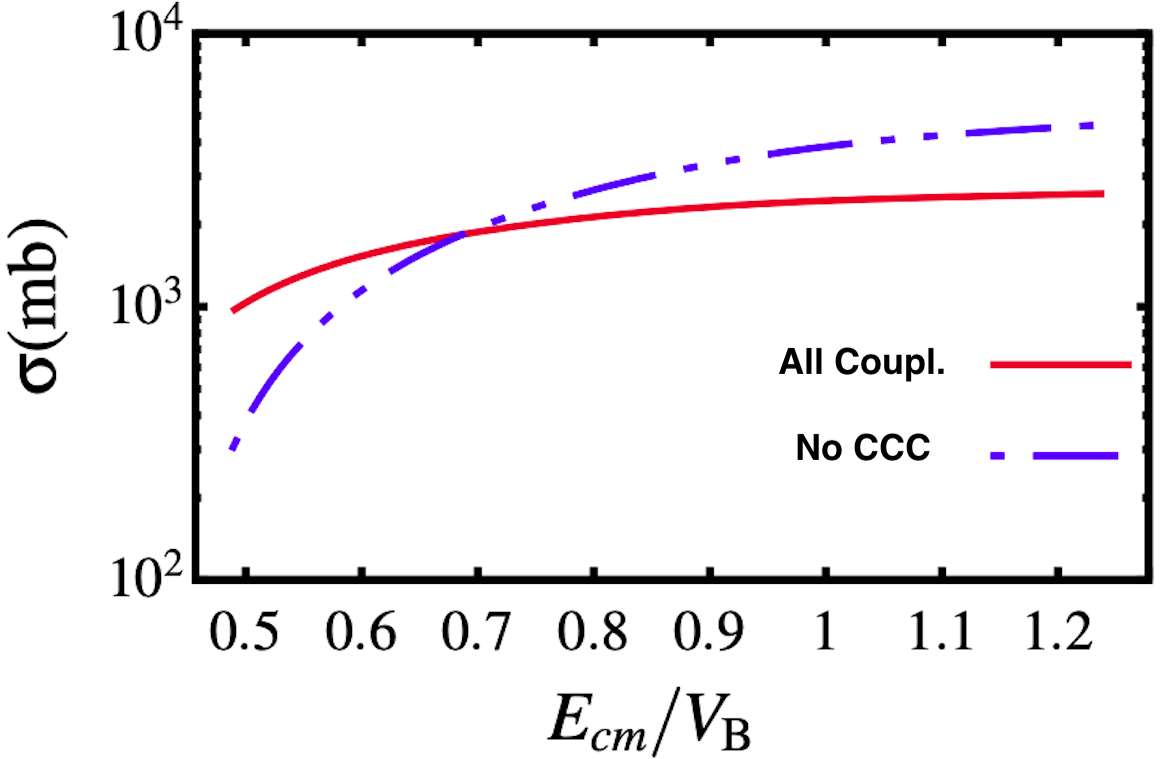}}
	\end{center}
	\caption{\label{f06}Breakup cross-sections plotted as functions of the incident energy scaled by
          the Coulomb barrier height $V_{\rm B}$, and obtained when all the different couplings are included in
          the coupling matrix elements ``All coupl.'' and when the continuum-continuum couplings are excluded
          from the couplings matrix elements ``No CCC''.}
\end{figure}

To better assess of the significance of the continuum-continuum couplings on the breakup
cross-section,
we plot in Fig. \ref{f06}, the breakup cross-sections in the presence and absence of the
continuum-continuum couplings.
  Observing that figure, it resorts that at deep sub-barrier energies ($E_{cm}/V_{\rm B}\le 0.7$), the continuum-continuum
  couplings serve to enhance the breakup cross-section, as the breakup cross-section in the presence of these couplings
  is larger than the breakup cross-section in the absence of these couplings. A similar trend is reported in
  Ref.\cite{MukeruJPG2018}, in the case of the breakup of the weakly-bound cluster system $^6$Li on the same target nucleus.
  At larger incident energies ($E_{cm}/V_{\rm B}> 0.7$), the continuum-continuum couplings  account for
  the suppression of the breakup cross-section.

  What can we learn from the enhancement of the breakup cross-section at sub-barrier
  energies?
Notice that, according to Refs.~\cite{Torres2018,Sunil2016}, the structure of 
the continuum, with existing resonances, may delay the breakup process. For 
instance, as stated in Ref.~\cite{Sunil2016},    
  \textit{``The near-target breakup is consistent with simulations which assume that 
  the populated continuum states have a short but finite mean life, delaying into 
  fragments."} The same authors further argue that:
  \textit{``Similar behavior should be expected from breakup occurring from very 
short-lived states, irrespective of whether the breakup is direct or 
triggered by transfer of nucleons."}

If it could also be verified that continuum-continuum couplings produce
similar effect, 
then, for energies above the Coulomb barrier, delaying the breakup would mean 
increasing the probability of the breakup to occur within the interaction region 
where nuclear forces are active to trigger absorption.
Therefore, at such incident energies higher than the barrier, this will lead to 
total fusion larger than the breakup cross-sections 
as verified in Fig.~\ref{f03}. 
In the same energy region, if the breakup becomes more prompt when the 
continuum-continuum couplings are removed, then the projectile fragments 
have enough time to survive absorption after the breakup, explaining the 
increase of the breakup cross-section over the total fusion cross-section
as observed in Fig.~\ref{f05}. 
However, until it is proven that continuum-continuum couplings in fact will
delay the breakup process, our assertion here remains speculative. 
A study in this regard could be important, by considering for example 
a dynamical time-dependent approach.

  At sub-barrier incident energies, having low kinetic energy, the projectile further 
  slows down due to the Coulomb repulsion. The dissipation of the projectile kinetic 
  energy can be exacerbated by including couplings among continuum, 
  since energies are required to excite these states. 
  As such, as an intermediate step, the projectile can be in a continuum 
  state, although asymptotically it ends up bound.
  However, on the outgoing trajectory, the projectile gains an initial kinetic energy
  as its leaves the target, considering the fact that in that case it is accelerated by the 
  projectile-target Coulomb force.
  On both incoming and outgoing projectile trajectories, the continuum-continuum 
  couplings play the same role in the breakup process.
  Again, assuming that the continuum-continuum couplings could delay the
    breakup, then such ``delay'' together with the projectile acceleration coming from the 
    projectile-target Coulomb interaction, can increase the probability of the projectile
    breakup on its outgoing trajectory away from the absorption region, i.e., out of reach 
  of nuclear forces. Consequently, the fusion cross-section would be reduced at the 
  expense of the breakup cross-section, with less amount of flux being removed from 
  the breakup channel to feed up the fusion channel.  The opposing effect of 
  the Coulomb interaction on both incoming and outgoing trajectories is also invoked in
  Ref.\cite{Yang2022} to explain the magnitude of the opening angle of the breakup
  fragments, considering that this angle distribution provides information on the 
  breakup location (as shown in Ref.~\cite{Simpson2016}).
  Therefore, we cautiously speculate that the enhancement of the breakup
  cross-section at sub-barrier incident energies 
  by the continuum-continuum couplings could be associated with the breakup 
  of the projectile occurring in the outgoing trajectory.
  Although it is not possible to unambiguously prove this point in this work, it
    could provide hints for further studies in this direction.
  More discussion on this 
  aspect can be found  in Refs.\cite{Yang2022,Zhang2023}.
  
  The argument ``on the outgoing trajectory the projectile can break up away from 
  the reach of the nuclear forces" can be substantiated by showing that, at sub-barrier 
  energies, the enhancement of the breakup cross-section is due 
  to its Coulomb breakup component. This inference is born out of the fact that on 
  the outgoing trajectory the nuclear breakup becomes increasingly less relevant as 
  the projectile moves away from the target nucleus. To shed more light on this, 
  let us analyze the Coulomb and nuclear breakup cross-sections.   

\begin{figure}[!!t]
	\begin{center}
			\resizebox{85mm}{!}{\includegraphics{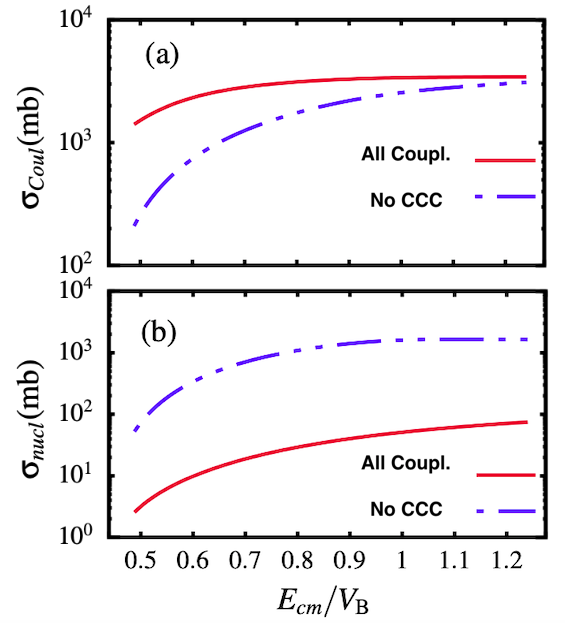}}
	\end{center}
	\caption{\label{f07}Coulomb breakup cross-sections [panel (a)] and nuclear breakup cross-sections
          [panel (b)]  plotted as functions of the incident energy scaled by
          the Coulomb barrier height $V_{\rm B}$, and obtained when the continuum-continuum couplings are
          included and excluded
          from the coupling matrix elements.}
\end{figure}

 Notice that the breakup cross-section that we have so far discussed is obtained by coherently including
  both Coulomb and nuclear interactions in the coupling matrix elements, this we also call 
  total (Coulomb + nuclear)
breakup cross-section.
The separation of the total breakup cross-section into its Coulomb and nuclear components is not a straightforward task,
and this work is not intended to perform such a task. To obtain the Coulomb and nuclear breakup 
cross-sections, we resort to the following approximate procedure. To calculate the Coulomb
  breakup cross-section, we removed all the core-target and neutron-target nuclear
  interactions from the coupling matrix elements, keeping only its monopole component in the elastic
  scattering channel. This potential (as in all other calculations), was obtained by folding the projectile
  ground-state density with the projectile-target potentials. In this case, the Coulomb breakup cross-section
  is affected by the absorption in the elastic scattering channel due to the imaginary component of
  this potential. Similarly, the nuclear breakup cross-sections were obtained by removing the core-target
  Coulomb potential
  from the coupling matrix elements, also keeping its monopole diagonal term in the elastic
  scattering channel. This approach, although approximate, has proven to yield the desired effect.
  
  The Coulomb and nuclear breakup cross-sections thus obtained are shown in Fig.\ref{f07}.
    Indeed, in that figure, we notice that at sub-barrier energies, the Coulomb breakup cross-section is strongly enhanced 
    by the continuum-continuum couplings [see panel (a)].
    As the incident energy increases, the enhancement strength decreases and the trend suggests that at
    higher incident energies, the continuum-continuum couplings would amount to a smaller effect on the Coulomb breakup
    cross-section. In fact, it has been shown that at higher incident energy, these couplings have very small suppression
    effect on the Coulomb breakup cross-section (see for example Ref.\cite{MukeruCPC2022}). In contrast,
    panel (b) of Fig.\ref{f07} shows that the nuclear breakup cross-section is strongly suppressed by continuum-continuum
    couplings at all the displayed incident energy regions. At energies above the barrier, the nuclear breakup cross-section
    is known to be strongly suppressed by these couplings compared to the Coulomb breakup cross-section. In fact,
     in  Ref.\cite{MukeruCPC2022}, this suppression is reported as one of the main reasons why the Coulomb
    breakup is more important than the nuclear breakup in reactions involving heavy targets.
    Comparing the effect of the continuum-continuum couplings on Coulomb and nuclear breakup cross-sections, it follows that
    the enhancement of the total breakup cross-section at sub-barrier incident energies due to these couplings comes
    exclusively from the Coulomb breakup.

\section{Conclusion}

In this paper, we have analyzed the breakup of the weakly-bound neutron-halo nucleus
$^{11}{\rm Be}$ 
on a lead target at the sub-barrier and around the barrier incident energies.
It is found that at deep 
sub-barrier energies, the breakup cross-section is dominant over the total fusion
cross-section, 
implying that it is the leading reaction channel at this incident energy range, as also
reported in the case of the 
proton-halo projectile $^8$B on the same target. The continuum-continuum couplings,
which are reported to enhance the breakup cross-section at sub-barrier energies, 
are found to be responsible for this feature. The enhancement of the breakup 
cross-section by these couplings, at sub-barrier energies, is found to come 
exclusively from its Coulomb component. Based on this, we are speculating that 
such enhancement of the Coulomb breakup cross-section by the continuum-continuum 
couplings could be associated with the breakup occurring on the outgoing trajectory, 
provided it is proven that these couplings delay the breakup process.
 
In summary, our study is confirming that the importance of the breakup channel 
over the total fusion channel, at energies below the Coulomb barrier, can also 
be extended to neutron-halo projectile on heavy targets.
In spite of the fact that a detailed study may be required, based on the available 
investigations, one can anticipate this conclusion as being a universal feature 
in the breakup of weakly bound projectiles on heavy targets.

\acknowledgments
\noindent
We acknowledge partial support from the Brazilian agency Conselho Nacional de Desenvolvimento 
Cient\'\i fico e Tecnológico [INCT-FNA Proc. 464898/2014-5 (BM, LT) and 
Proc. 304469/2019-0(LT)].

\end{document}